\documentclass[aps,twocolumn,showpacs,preprintnumbers]{revtex4}

\newcommand{\bra}[1]{\langle #1|}
\newcommand{\ket}[1]{|#1\rangle}

\usepackage{graphicx}
\usepackage{dcolumn}
\usepackage{bm}
\usepackage{amsmath}

\normalsize

\begin{document}
\bibliographystyle{apsrev}

\title{Probing Decoherence with Electromagnetically Induced
Transparency in Superconductive Quantum Circuits}

\author{K.~V.~R.~M. Murali, D.~S. Crankshaw, and T.~P. Orlando}
\affiliation{Department of Electrical Engineering and Computer Science \\
Massachusetts Institute of Technology, Cambridge MA 02139}
\author{Z. Dutton}
\affiliation{National Institute of Standards \& Technology,
Electron and Optical Division, Gaithersburg MD 20899-8410}
\author{W.~D. Oliver}
\affiliation{MIT Lincoln Laboratory, 244 Wood Street, Lexington,
MA 02420}

\date{\today}

\begin{abstract}
Superconductive quantum circuits (SQCs) comprise quantized energy
levels that may be coupled {\em via} microwave electromagnetic
fields. Described in this way, one may draw a close analogy to
atoms with internal (electronic) levels coupled by laser light
fields. In this Letter, we present a superconductive analog to
electromagnetically induced transparency (S-EIT) that utilizes SQC
designs of present day experimental consideration. We discuss how
S-EIT can be used to establish macroscopic coherence in such
systems and, thereby, utilized as a sensitive probe of
decoherence.


\end{abstract}
\pacs{42.50.Gy}
\maketitle

\date{\today}

Superconductive quantum circuits (SQCs) comprising mesoscopic
Josephson junctions can exhibit quantum coherence amongst their
macroscopically large degrees of freedom~\cite{Leggett85a}. They
exhibit quantized flux and/or charge states depending on their
fabrication parameters, and the resultant quantized energy levels
are analogous to the quantized internal levels of an atom.
Spectroscopy, Rabi oscillation, and Ramsey interferometry
experiments have demonstrated that SQCs behave as ``artificial
atoms'' under carefully controlled
conditions~\cite{Nakamura99a,Friedman00a,Wal00a,Vion02a,Yu02a,Martinis02a,Chiorescu03a,Pashkin03a}.
This Letter extends the SQC-atom analogy to another quantum
optical effect associated with atoms: 
electromagnetically induced transparency
(EIT)~\cite{Boller91a,Harris97a}. We propose the demonstration of
microwave transparency using a superconductive analog to EIT
(denoted S-EIT) in a superconductive circuit exhibiting two
meta-stable states (e.g., a qubit) and a third, shorter-lived
state (e.g., the readout state). We show that driving coherent
microwave transitions between the qubit states and the readout
state is a demonstration of S-EIT. We further propose a means to
use S-EIT to experimentally probe the qubit decoherence rate in a
sensitive manner.  The philosophy is similar to that in
Ref.~\onlinecite{Ruostekoski99a}, where it was proposed to use EIT
to measure phase diffusion in atomic Bose-Einstein condensates.

\begin{figure}
\includegraphics{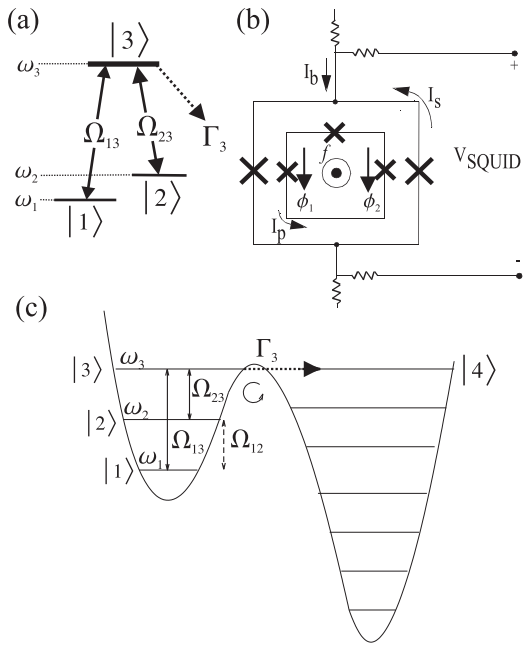}
\caption{\label{fig:diagram} \textbf{(a)} Energy level diagram of
a three-level $\Lambda$ system. EIT can occur in atoms possessing
two long-lived states $\ket{1},\ket{2}$, each of which is coupled
\em{via }\em resonant laser light fields to a radiatively decaying
state $\ket{3}$.  State $\ket{3}$ can feed back into
$\ket{1},\ket{2}$ and/or decay into levels outside the $\Lambda$
configuration. \textbf{(b)} Circuit schematic of the PC qubit and
its readout SQUID. \textbf{(c)} Schematic energy level diagram for
a three-level superconducting quantum circuit. For our parameters
we calculate $\omega_2-\omega_1=(2 \pi)~36~\mathrm{GHz}$ and
$\omega_3-\omega_2=(2 \pi)~32~\mathrm{GHz}$.  The simulated matrix
elements are $\bra{p}\mathrm{sin}(2 \pi f+2 \phi_m)\ket{q}$ for
$(p,q)=$ (1,2),(2,3), and (1,3) are, respectively, 0.0704, -0.125,
0.0158.}
\end{figure}

The three-level $\Lambda$ system illustrated in
Fig.~\ref{fig:diagram}a is a standard energy level structure
utilized in EIT~\cite{Boller91a,Harris97a}. It comprises two
meta-stable states $\ket{1}$ and $\ket{2}$, each of which may be
coupled to a third excited state $\ket{3}$. In atoms, the
meta-stable states are typically hyperfine or Zeeman levels, while
state $\ket{3}$ is an excited electronic state that may
spontaneously decay at a relatively fast rate $\Gamma_3$.  In an
atomic EIT scheme, a resonant ``probe'' laser couples the $\ket{1}
\leftrightarrow \ket{3}$ transition, and a ``control'' laser
couples the $\ket{2} \leftrightarrow \ket{3}$ transition. The
transition coupling strengths are characterized by their Rabi
frequencies $\Omega_{j3}\equiv-\mathbf{d}_{j3} \cdot
\mathbf{E}_{j3}$ for $j=1,2$ respectively, where $\mathbf{d}_{j3}$
are the dipole matrix elements and $\mathbf{E}_{j3}$ are the
slowly varying envelopes of the electric fields. For particular
Rabi frequencies $\Omega_{j3}$, the probe and control fields are
effectively decoupled from the atoms by a destructive quantum
interference between the states of the two driven transitions. The
result is probe and control field
transparency~\cite{Boller91a,Harris97a}. In more recent
experiments, ultra-slow light propagation due to EIT-based
refractive index modifications in atomic clouds have also been
demonstrated~\cite{Hau99a,Kash99a,Budker99a}.

SQCs have also been demonstrated to exhibit $\Lambda$-like energy
level structures~\cite{Martinis02a,Segall03a,Crankshaw03a,Yu03a}.
One example is the persistent-current (PC) qubit, a
superconductive loop interrupted by two Josephson junctions of
equal size and a third junction scaled smaller in area by the
factor $\alpha <1$
(Fig.~\ref{fig:diagram}b)~\cite{Orlando99a,Mooij99a}. Its dynamics
are described by the Hamiltonian
\begin{multline}
 \label{eq:pcq}
 {\cal{H}}_{pc} = \frac{1}{2}C
 \left( \frac{\Phi_{0}}{2\pi} \right)^{2}
 (\dot{\varphi}^{2}_{p} + (1 + 2 \alpha) \dot{\varphi}^{2}_{m}) \\
  + E_{j} \left[2 + \alpha - 2 \cos\varphi_{p} \cos\varphi_{m} -
 \alpha \cos(2 \pi f + 2 \varphi_{m})\right],
\end{multline}
in which $C$ is the capacitance of the larger junctions,
$\varphi_{p,m} \equiv (\varphi_{1} \pm \varphi_{2})/2$,
$\varphi_{i}$ is the gauge-invariant phase across the larger
junctions $i=\{ 1,2 \}$, $E_J$ is the Josephson coupling energy,
and $f$ is the magnetic flux through the loop in units of the flux
quantum $\Phi_{0}$. The second term in Eq.~(\ref{eq:pcq}) defines
the magnetic-flux-dependent qubit potential landscape. For flux
biases near one-half of a flux quantum, $f \approx 1/2$, the
potential may be approximated by a double-well potential, with
each well corresponding to a distinct, stable, classical state of
the electric current, i.e., left or right circulation through the
loop. In turn, each current state has a net magnetization of
opposite direction that is measurable using a dc
SQUID~\cite{Orlando99a}. As a quantum object, the potential wells
exhibit quantized energy levels corresponding to the quantum
states of the macroscopic circulating
current~\cite{Segall03a,Crankshaw03a}. These levels may be coupled
using microwave radiation~\cite{Wal00a,Yu03a}, and their quantum
coherence has been experimentally
demonstrated~\cite{Chiorescu03a}. Note that for this system, the
terms ``population'' and ``occupation probability'' are used
synonymously.

Tuning the flux bias away from $f=1/2$ results in the asymmetric
double-well potential illustrated in Fig.~\ref{fig:diagram}c. The
three states in the left well constitute the superconductive
analog to the atomic $\Lambda$ system. States $\ket{1}$ and
$\ket{2}$ are ``meta-stable states,'' with a tunneling and
coherence time much longer than the excited ``readout'' state
$\ket{3}$. State $\ket{3}$ has weakly-coupled intra-well
transitions, but has a strong inter-well transition when tuned on
resonance with state
$\ket{4}$~\cite{Segall03a,Crankshaw03a,Yu03a}. Using tight-binding
models with experimental PC qubit
parameters~\cite{Orlando99a,Yu03a,Crankshaw03a} at a flux bias
f=0.5041, we estimate the tunneling times from states $\ket{1}$,
$\ket{2}$, and $\ket{3}$ to the right well are $1/\Gamma_{1}
\approx$ 1~ms, $1/\Gamma_{2} \approx$ 1~$\mu$s, and $1/\Gamma_{3}
\approx$ 1~ns respectively. Thus, a particle reaching state
$\ket{3}$ will tend to tunnel quickly to state $\ket{4}$, and this
event results in a switching of the circulating current that may
be detected using a fast-measurement scheme. Alternatively, one
may detune states $\ket{3}$ and $\ket{4}$, and then apply a
resonant $\pi$-pulse to transfer the population from state
$\ket{3}$ to $\ket{4}$; since the states are now off-resonance,
the relaxation rate back to the left well is reduced and a slower
detection scheme may be used. We note that a single-junction
qubit~\cite{Martinis02a} shares this property, since the right
well in Fig.~\ref{fig:diagram}c is effectively replaced by a
quasi-continuum of states, and transitions out of the left well
will not return.

Transitions between the quantized levels are driven by resonant
microwave-frequency magnetic fields.  Assuming the Rabi
frequencies $\Omega_{ij}$ to be much smaller than all level
spacings $|\omega_{kl}| \equiv |\omega_k - \omega_l|$, the
system-field interaction may be written within the rotating wave
approximation (RWA)\cite{Scully97a},
\begin{align}
 {\cal{H}}^{(\mathrm{RWA})}_{int} = \frac{\hbar}{2} \left[\begin{matrix}
         0 & \Omega_{12}^* & \Omega_{13}^* \\
         \Omega_{12} & 0 & \Omega_{23}^* \\
      \Omega_{13}& \Omega_{23} & -i  \Gamma_3
         \end{matrix}\right],
 \label{eq:interactionframe1}
\end{align}
in which the decay from state $\ket{3}$ is treated
phenomenologically as a non-Hermitian matrix
element~\cite{Scully97a,Dutton02a}.  For small microwave
perturbations of amplitude $f_{\Delta}$, the associated Rabi
frequencies are given by $\Omega_{pq}= f_{\Delta} \bra{p} \sin{(2
\pi f + 2 \phi_{m})} \ket{q}$; numerical simulations of the matrix
elements using PC qubit parameters are consistent with recent
experimental results (see caption
Fig.~\ref{fig:diagram}c)~\cite{Segall03a,Crankshaw03a,Yu03a}. In
general, all three intra-well transitions are allowed in SQCs. For
example, consider states $\ket{1}$ and $\ket{2}$ to be a qubit
that is prepared in an arbitrary superposition state
$\ket{\Psi}=c_{1}\ket{1} + c_{2}\ket{2}$ by temporarily driving
the $\Omega_{12}$ transition. Then, by applying a probe field, the
population of state $\ket{1}$ may be read out through a transition
to state $\ket{3}$ followed by a rapid escape to the right
well~\cite{Martinis02a}. In this case, both the preparation and
readout transitions were allowed and absorptive.

One may achieve S-EIT in a superconductive $\Lambda$ system that
is prepared in state $\ket{\Psi}=c_{1}\ket{1} + c_{2}\ket{2}$, by
simultaneously and solely applying the microwave fields
$\Omega_{13}$ and $\Omega_{23}$ such that
\begin{align}
 \frac{\Omega_{13}}{\Omega_{23}}=-\frac{c_2}{c_1}.
 \label{eq:darkState}
\end{align}
Under this condition (with $\Omega_{12}=0$), the state
$\ket{\Psi}$ is an eigenstate of
${\cal{H}}^{(\mathrm{RWA})}_{int}$ in
Eq.~(\ref{eq:interactionframe1}) with eigenvalue zero, and the SQC
becomes transparent to the microwave fields. As in conventional
EIT, the amplitudes for the two absorption transitions into
$\ket{3}$ have equal and opposite probability amplitudes, leading
to a destructive quantum interference. Thus, in the absence of
decoherence, preparing the qubit in state $\ket{\Psi}$ with an
ideal preparation ($\Omega_{12}$) field and subsequently applying
ideal probe ($\Omega_{13}$) and control ($\Omega_{23}$) microwave
fields which satisfy Eq.~(\ref{eq:darkState}) would result in no
population loss through the readout state $\ket{3}$. In this way,
S-EIT would confirm, without disturbing the system, that we had
indeed prepared the qubit in the desired state.

In a practical SQC, there will be decoherence of the state
$\ket{\Psi}$, and this must be measured, characterized, and
minimized for quantum information applications. S-EIT is one such
sensitive decoherence probe, since deviations in the amplitude
and/or relative phase of the complex coefficients $c_i$ from the
condition established in Eq.~(\ref{eq:darkState}) result in a
small probability $|(c_1 \Omega_{13}+c_2 \Omega_{23})/\Omega|^2$
of the SQC being driven into the readout state $\ket{3}$ on a time
scale $\sim \Gamma_3/\Omega^2$. In general, there are two
categories of decoherence: {\it loss} and {\it dephasing}. {\it
Loss} refers to population losses from the metastable states
$\ket{1},\ket{2}$, and it is present in an SQC due to, for
example, the finite loss rate of level $\ket{2}$, $\Gamma_2 \sim
1/\mu\mathrm{s}= (2 \pi)~0.2~\mathrm{MHz}$. {\it Dephasing} refers
to interactions of the SQC with other degrees of the freedom in
the system that cause the relative phase between $c_1$ and $c_2$
to diffuse.
The incorporation of dephasing is facilitated by the use of a
density matrix formalism.

We describe the system with a $3 \times 3$ density matrix with
diagonal elements $\rho_{ii}$ describing the populations, and
$\rho_{ij},~i \not= j$ describing the coherences between levels.
In the presence of the EIT fields $\Omega_{13}$ and $\Omega_{23}$
with no direct coupling ($\Omega_{12}=0$), the Bloch equations
govern the evolution of the density matrix~\cite{Scully97a}:
\begin{align}
 \label{eq:OBEdephase11} \dot{\rho}_{11} &= - \Gamma_1 \rho_{11}
 -\frac{i}{2} \Omega_{13}^* \rho_{31} + \frac{i}{2} \Omega_{13} \rho_{13},  \\
 \label{eq:OBEdephase22} \dot{\rho}_{22} &=  - \Gamma_2 \rho_{22}
 - \frac{i}{2}
 \Omega_{23}^* \rho_{32} + \frac{i}{2} \Omega_{23} \rho_{23},  \\
 \label{eq:OBEdephase33} \dot{\rho}_{33} &= - \Gamma_3 \rho_{33}
 + \frac{i}{2} \Omega_{13}^* \rho_{31} - \frac{i}{2} \Omega_{13}
 \rho_{13} \nonumber \\ &  + \frac{i}{2}
 \Omega_{23}^* \rho_{32} - \frac{i}{2} \Omega_{23} \rho_{23},  \\
 \label{eq:OBEdephase12} \dot{\rho}_{12} &=
 -\gamma_{12}\rho_{12}-\frac{i}{2}\Omega_{13}^*\rho_{32}
 +\frac{i}{2}\Omega_{23}\rho_{13},  \\
 \label{eq:OBEdephase13} \dot{\rho}_{13} &=
 - \gamma_{13}\rho_{13}+\frac{i}{2}\Omega_{13}^*(\rho_{11}-\rho_{33})
 +\frac{i}{2}\Omega_{23}^*\rho_{12},  \\
 \label{eq:OBEdephase23} \dot{\rho}_{23} &=
 -\gamma_{23}\rho_{23}+\frac{i}{2}\Omega_{23}^*(\rho_{22}-\rho_{33})
 +\frac{i}{2}\Omega_{13}^*\rho_{21}.
\end{align}

\noindent The remaining three elements' equations are determined
by $\rho_{ij}^*=\rho_{ji}$.  The decoherence rates
$\gamma_{ij}=(\Gamma_i+\Gamma_j)/2 +
\gamma_{ij}^{\mathrm{(deph)}}$ include both loss and dephasing
contributions. We concentrate on the regime in which the readout
state escape rate $\Gamma_3=1 \text{
ns}^{-1}=(2\pi)~130~\mathrm{MHz}$ dominates all other loss and
dephasing rates, thus $\gamma_{13} \approx \gamma_{23} \approx
\Gamma_3/2$. Furthermore, we ignore the meta-stable state losses
$\Gamma_1,\Gamma_2$ relative to the dephasing
$\gamma_{12}^{\mathrm{(deph)}}$ and set $\gamma_{12} \approx
\gamma_{12}^{\mathrm{(deph)}}$. Theoretical estimates of dephasing
rates, such as $\gamma_{12}^{\mathrm{(deph)}}$, in multi-level
systems were recently obtained in Ref.~\onlinecite{Burkhard03a}.

\begin{figure}
\includegraphics{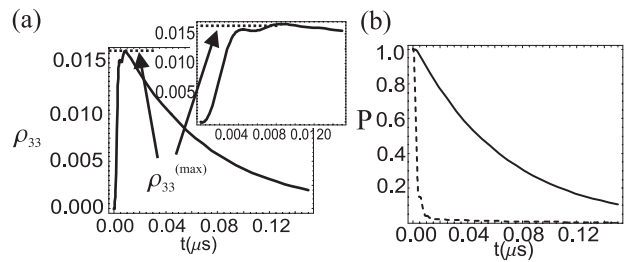}
\caption{\label{fig:dephaseExample}\textbf{(a)} The population
(occupation probability) $\rho_{33}$ as a function of time with
EIT fields $\Omega_{13}=\Omega_{23}=(2 \pi)~150~\mathrm{MHz}$
applied to an initial dark state $\rho_{11}=\rho_{22}=0.5$ and
$\rho_{12}=-0.5$, and with a dephasing rate $\gamma_{12}=(2
\pi)~5~\mathrm{MHz}$. The inset shows the same curve zoomed in on
the early times. Under conditions of good EIT, we see a rapid
initial rise to some plateau, followed by a much slower decay.
\textbf{(b)} The total population $P(t)$ remaining in the system
versus time for the same simulation (solid curve). For comparison,
the dashed curve shows the population for the out of phase case
$\rho_{11}=\rho_{22}=0.5$ and $\rho_{12}=0.5$ discussed in the
text.}
\end{figure}

We illustrate an S-EIT decoherence probe example by applying EIT
fields $\Omega_{13}=\Omega_{23}=(2 \pi)~150~\mathrm{MHz}$ to the
dark state $\ket{\Psi}=(\ket{1} - \ket{2})/\sqrt{2}$ with a
dephasing rate $\gamma_{12}=(2 \pi)~5~\mathrm{MHz}$ and
numerically integrating
Eqs.~(\ref{eq:OBEdephase11})-(\ref{eq:OBEdephase23}). The
dephasing $\gamma_{12}$ causes a small population transfer to the
excited state $\rho_{33}$ (Fig.~\ref{fig:dephaseExample}a). The
excited state population initially exhibits a rapid rise (see
inset Fig.~\ref{fig:dephaseExample}a) with transitory
oscillations, reaching its maximum value
$\rho_{33}^{(\mathrm{max})}$ within about 4~ns. This is followed
by a smooth decay with a $1/e$ time of about 80~ns. The solid
curve in Fig.~\ref{fig:dephaseExample}b traces the
total population $P=\rho_{11}+\rho_{22}+\rho_{33}$ remaining in
the system as a function of time. When the excited state maximum
$\rho_{33}^{(\mathrm{max})}$ is reached,
the total remaining population is $P(4 \text{ ns})=0.973$. In
contrast, the dashed line in Fig.~\ref{fig:dephaseExample}b
illustrates the rapid population loss expected when the same
fields are applied to the state
$\ket{\Psi}=(\ket{1}+\ket{2})/\sqrt{2}$ [$\pi$ out of phase with
the dark state in Eq.~(\ref{eq:darkState})]. In the absence of
S-EIT quantum interference, the entire population is lost on a
time scale $\Gamma_3/\Omega^2 \sim 1~\mathrm{ns}$. The general
behavior presented in Figs.~\ref{fig:dephaseExample}a
and~\ref{fig:dephaseExample}b is observed over a wide parameter
regime of experimental interest.

We now use Eqs.~(\ref{eq:OBEdephase11})-(\ref{eq:OBEdephase23}) to
show how measuring the slow population loss in S-EIT can be used
to extract the decoherence rate $\gamma_{12}$. The elements
$\rho_{33}$, $\rho_{13}$, and $\rho_{23}$ in
Eqs.~\ref{eq:OBEdephase33},~\ref{eq:OBEdephase13},
and~\ref{eq:OBEdephase23} are damped at a rapid rate $\sim
\Gamma_3$, allowing their adiabatic
elimination~\cite{Javanainen95a,Dutton02a}; we solve for their
quasi-steady state values by setting
$\dot{\rho}_{33}=\dot{\rho}_{13}=\dot{\rho}_{23}=0$. This
approximation is accurate once initial transients have passed and
the plateau value $\rho_{33}^{(\mathrm{max})}$ has been reached.
Using these results in Eq.~(\ref{eq:OBEdephase12}) yields an
equation for $\dot{\rho}_{12}$ with a strong damping term
$\Omega^2/\Gamma_3$, and it too can be solved for its quasi-steady
state value.  In the limit $\gamma_{12} \Gamma_3/\Omega^2 \ll 1$
we get~\cite{Murali03a}
\begin{align}
 \label{eq:rho12ss}
  \rho_{12}(t) \approx
   -\frac{\Omega_{13}\Omega_{23}}{\Omega^2}\bigg(1 - \frac{2
   \gamma_{12} \Gamma_3}{\Omega^2}\bigg)
   \big( \rho_{11}(t)+\rho_{22}(t) \big).
\end{align}

\noindent The ratio $2\gamma_{12} \Gamma_3/\Omega^2$ represents
the small fractional deviation of $\rho_{12}$ from its dark state
value. There is a competition between the ``preparation rate''
$\Omega^2/\Gamma_3$ (which constantly acts to drive the system
into the dark state) and the decoherence rate $\gamma_{12}$ (which
attempts to drive it back out).

We now consider the conditions under which one can use S-EIT to
estimate the decoherence rate with little population loss in the
system. Eqs.~(\ref{eq:OBEdephase11}), ~(\ref{eq:OBEdephase22}),
and ~(\ref{eq:rho12ss}) reveal that deviations from the dark state
cause population loss through $\ket{3}$ at a rate $R=2 \gamma_{12}
(\Omega_{13}^2 \Omega_{23}^2/\Omega^4)$; this ultimately leads to
the exponential decay of $P$ seen in
Fig.~\ref{fig:dephaseExample}b. By assumption, population escapes
the system only through the decay term $-\rho_{33} \Gamma_3$ in
Eq.~(\ref{eq:OBEdephase33}), yielding $\rho_{33}=(R/\Gamma_3) P$.
At the early time when $\rho_{33}^{\mathrm{(max)}}$ has been
reached, the population $P$ remains close to unity, and the
excited state population reaches
\begin{align}
 \label{eq:P3max}
 \rho_{33}^{\mathrm{(max)}} \approx 2
   \frac{\Omega_{13}^2 \Omega_{23}^2}{\Omega^4}
   \frac{\gamma_{12}}{\Gamma_3}.
\end{align}

\noindent The time $T_{ss}$ to reach $\rho_{33}^{\mathrm{(max)}}$
is generally the smaller of the preparation time $\sim
\Gamma_3/\Omega^2$ and the inverse of the decay rate $1/\Gamma_3$.
At this time, the total population loss will be $\sim
T_{ss}\rho_{33}^{\mathrm{(max)}} \sim (2
\Omega_{13}^2\Omega_{23}^2/\Omega^4)\,\mathrm{Max}(\gamma_{12}/\Gamma_3,\gamma_{12}
\Gamma_3/\Omega^2)$. So long as the loss during this initial
transient time is small, the population will follow a simple
exponential decay $P(t)= \exp(-\rho_{33}^{\text{(max)}} \Gamma_3
t)$, and the dephasing rate $\gamma_{12}$ can be easily extracted.
To keep this loss small, we require both ratios in the
$\mathrm{Max}(\cdots)$ argument to be small ($\Omega \gg \sqrt{2
\gamma_{12} \Gamma_3}$ and $\Gamma_3 \gg \gamma_{12}$) in order to
use this approach to accurately estimate the decoherence rate
while causing little loss from the system. Since $\Omega$ is
experimentally controllable, it can be chosen to satisfy the first
constraint. If $\Gamma_3$ is comparable or smaller than
$\gamma_{12}$, then S-EIT remains a decoherence probe, although
the strong damping assumption leading to Eq.~(\ref{eq:rho12ss}) no
longer holds and so the analysis is different.

\begin{figure}
\includegraphics{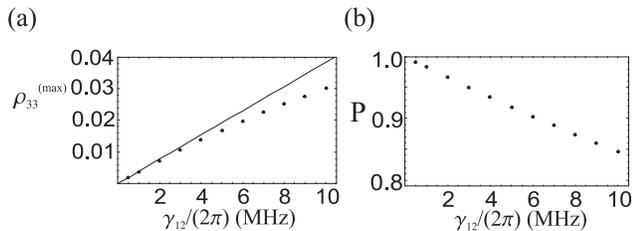}
\caption{\label{fig:dephaseSeries}\textbf{(a)} The maximum plateau
value $\rho_{33}^{(\mathrm{max})}$ for different $\gamma_{12}$
(circles).  The solid curve shows the prediction (\ref{eq:P3max}).
\textbf{(b)} The remaining population
$P=\rho_{11}+\rho_{22}+\rho_{33}$ at the time the plateau is
reached for the cases in (a). }
\end{figure}

We have performed a series of numerical simulations, varying
$\gamma_{12}$ to test the validity of the above approach. The
results are presented in Fig.~\ref{fig:dephaseSeries}.
Fig.~\ref{fig:dephaseSeries}a indicates
$\rho_{33}^{(\mathrm{max})}$ versus $\gamma_{12}$ and compares the
results with the analytic estimate [Eq.~(\ref{eq:P3max})]. The
agreement is quite good for $\gamma_{12}<(2 \pi)~4~\mathrm{MHz}$,
which corresponds to the $2 \gamma_{12} \Gamma_3/\Omega^2<0.056$.
Higher dephasing rates compete more with the preparation rate,
making the adiabatic elimination approach less valid; this leads
to deviations from our analytic prediction [Eq.~(\ref{eq:P3max})].
In such cases, one observes a significant loss by the time
$\rho_{33}^{(\mathrm{max})}$ is reached, as illustrated in
Fig.~\ref{fig:dephaseSeries}b

The RWA invoked in the calculation ignores far off-resonance
couplings induced by other applied fields at other transitions
({\it e.g.}, in our case, the $\Omega_{23}$ field drives
$\ket{1}\leftrightarrow\ket{2}$ at $\Delta \sim 4~\mathrm{GHz}$
off resonance).  We have performed calculations including all such
off-resonant couplings. Generally, they lead to small shifts of
the energy levels (analogous to A.C. Stark shifts) and loss rates.
These losses scale as $\Gamma_3(\Omega^2/\Delta^2)$ and
$\Omega^4/\Gamma_3^2 \Delta$, putting a limit on the field
strengths $\Omega^2$ which can be used for a given $\Delta$. For
the parameters considered here, we found shifts of $(2
\pi)~8~\mathrm{MHz}$ and loss rates totalling $(2
\pi)~19~\mathrm{kHz}$; this should not effect measurements in the
regime $\gamma_{12} \gg (2 \pi)~19~\mathrm{kHz}$~\cite{Murali03a}.

We have proposed using the superconductive analog to EIT (S-EIT)
to demonstrate macroscopic quantum interference in superconductive
quantum circuits. S-EIT provides an accurate and sensitive means
to probe the accuracy and phase coherence of qubit preparation,
and we have calculated analytic expressions for the field
strengths required for this purpose.

This work was supported in part by the AFOSR grant
No.~F49620-01-1-0457 under the Department of Defense University
Research Initiative in Nanotechnology (DURINT). The work at
Lincoln Laboratory was sponsored by the AFOSR under Air Force
Contract No.~F19628-00-C-0002.

\end{document}